\def\be{\begin{equation}}
\def\ee{\end{equation}}
\def\ba{\begin{array}}
\def\ea{\end{array}}

\documentclass[prl,showpacs,twocolumn,amsmath]{revtex4}
\usepackage{mathrsfs}
\usepackage{amssymb}
\usepackage{pifont}
\usepackage{epsfig,subfigure,dsfont,amsthm,amsbsy,mathrsfs,amscd}
\def\qed{\leavevmode\unskip\penalty9999 \hbox{}\nobreak\hfill
     \quad\hbox{\leavevmode  \hbox to.77778em{%
               \hfil\vrule   \vbox to.675em%
               {\hrule width.6em\vfil\hrule}\vrule\hfil}}
     \par\vskip3pt}

\newtheorem{theorem}{Theorem}
\newtheorem{corollary}{Corollary}

\input amssym.def
\begin{document}
\title{\large\bf Trade-off relations of $l_1$-norm coherence for multipartite systems}
\author{Zhengmin Jiang$^{1}$, Tinggui Zhang$^{1, \dag}$, Xiaofen Huang$^{1}$, Shao-Ming Fei$^{2,3}$}
\affiliation{ $^{1}$School of Mathematics and Statistics, Hainan
Normal University, Haikou, 571158, China\\
\footnotesize\small $^{2}$ School of Mathematical Sciences, Capital
Normal University, Beijing
100048,China\\
 \footnotesize \small
$^{3}$ Max-Planck-Institute for Mathematics in
the Sciences, Leipzig 04103,Germany \\
\footnotesize\small $^{\dag}$ Correspondence to tinggui333@163.com}
\date{}
\bigskip

\begin{abstract}
We study the trade-off relations given by the $l_{1}$-norm coherence
of general multipartite states. Explicit trade-off inequalities are
derived with lower bounds given by the coherence of either bipartite
or multipartite reduced density matrices. In particular, for pure
three-qubit states, it is explicitly shown that the trade-off
inequality is lower bounded by the three tangle of quantum entanglement.
\end{abstract}

 \maketitle

\bigskip
\section{Introduction } As one of the central concepts in quantum mechanics
which distinguish quantum from classical physics, coherence plays
 a significant role in many quantum phenomena such as the phase space distributions in quantum optics \cite{rjgl} and higher order correlation
functions \cite{esud}. It is highly desirable to
precisely quantify the usefulness of coherence as a resource.
In the classical work of Baumgratz, Cramer and Plenio \cite{tbmc}, this was
achieved by defining the key concepts such as incoherent states,
maximally coherent states and incoherent operations. Rapid
developments have been made since then on the fundamental theory of quantum coherence
and its applications \cite{asga1,mhxh}.

A successful and secure quantum network relies on quantum
correlations distributed among the subsystems \cite{hjki}. The
so-called monogamy relation of the distribution of quantum resources
characterizes such correlation distributions. In \cite{vcjk} it was
first time shown that the concurrence of the reduced states $\rho_{AB}$
and $\rho_{AC}$ of an arbitrary tripartite state $\rho_{ABC}$
satisfies the Coffman-Kundu-Wootters relation, where concurrence is an entanglement measure.
 Monogamy relations have been investigated for quantum entanglement
\cite{tjfv,xzsf,liuf,yggg}, quantum discord \cite{asga,yknz},
quantum streering \cite{mdam}, Bell nonlocality
\cite{vsng,btfv,pktp,qhfs,scmj}, indistinguishability \cite{mkdk},
other nonclassical correlations \cite{scll,zjsf} and quantum
coherence \cite{crmp,cpsd,bkll}.

Distributions of different quantum resources have been also studied,
such as the fundamental monogamy relation between contextually and
nonlocality \cite{pkac}, Bell nonlocality and three tangle for
three-qubit states \cite{ppam}, the internal entanglement and
external correlations \cite{scam1,scam2}. More recently, the
trade-off relations for Bell inequality violations in qubit networks
\cite{rrpm}, for quantum steering in distributed scenario
\cite{arss}, for state-dependent error-disturbance in sequential
measurements \cite{ymzm}, and for the entanglement cost and
classical communication complexity of causal order structure of a
protocol in distributed quantum information processing \cite{ewas}
have been investigated.

The distribution of quantum coherence in multipartite systems based
on relative entropy is given in \cite{crmp,cpsd,bkll} with nice
geometrical intuition, although the relative entropy is not easily
calculated. The $N$-partite monogamy of coherence is given
by defining $M=\sum_{n=2}^N C_{1:n}-C_{1:2\cdots N}$, where $C_{1:n}$ is the
intrinsic coherence between the partitions 1 and $n$ \cite{crmp}. For
$M\leq0$, i.e, $C_{1:2\cdots N}\geq\sum_{n=2}^N C_{1:n}$, one obtains a
monogamy relation (e.g. for the GHZ states). If $M>0$, i.e, $C_{1:2\cdots N}<\sum_{n=2}^N C_{1:n}$, one has a polygamous
relation (e.g. for the W states).

In \cite{tbmc} two different measures of coherence, the relative entropy of coherence $C_{r}$ and the $l_{1}$ norm of coherence $C_{l_1}$, have been proposed.
$C_{r}$ is an entropic measure, while $C_{l_{1}}$ is a geometric (distance) measure. Both $C_{r}$ and $C_{l_1}$
satisfy the strong monotonicity for all states, and the corresponding quantum resources theories have been
rigorously established \cite{tbmc}. Some relations between $C_{r}$ and $C_{l_1}$ have been also studied in \cite{srpp}.

For any $d$-dimensional quantum state $\rho$, one has
$C_{r}(\rho)\leq log(d)$, where the upper bound is attained for
maximally coherent states,
$|\varphi\rangle=\frac{1}{\sqrt{d}}\sum_{i=1}^{d}|i\rangle$
\cite{tbmc,awaa}. For bipartite states $\rho_{AB}$, its correlated
coherence $C_{cc}(\rho_{AB})$ is defined by
$C_{cc}(\rho_{AB})=C_{l_{1}}(\rho_{AB})- C_{l_{1}}(\rho_{A})-C_{l_{1}}(\rho_{B})$,
with $\rho_{A}$ and $\rho_{B}$ the reduced density matrices of the
subsystems. $C_{cc}(\rho_{AB})$ is always nonnegative \cite{awwa}.
Namely, $C_r(\rho_{AB})\geq C_r(\rho_{A})+C_r(\rho_{B})$, which
gives a kind of trade-off relations among the bipartite coherence
and the coherence of the subsystems \cite{awaa}. $C_{f}(\rho_{AB})\geq C_{f}(\rho_{A})+C_{f}(\rho_{B})$ is given in \cite{ymzz}, $C_{f}(\rho)$ is a convex roof coherence measure, and defined as $C_{f}(\rho)=\inf_{\{p_{i},|\varphi_{i}\rangle\}}\sum_{i}p_{i}C_{f}(|\varphi_{i}\rangle)$ with $\rho=\sum_{i}p_{i}|\varphi_{i}\rangle\langle\varphi_{i}|$. For tripartite
states $\rho_{ABC}$, \cite{ewaa} has been discussed whether a
similar trade-off relation like $C_r(\rho_{ABC})\geq
C_r(\rho_{AB})+C_r(\rho_{AC})$ holds. Unfortunately, this conjecture
is invalid. An interesting and important question one would ask is
then what trade-off relations hold among the tripartite or
multipartite coherence and the coherence of the reduced subsystems.

In this paper, we investigate the distribution of quantum
coherence in multi-qubit systems by using the easily calculated
$l_{1}$-norm of quantum coherence \cite{tbmc}. We derive explicit
trade-off inequalities lower bounded by the coherence of either
bipartite or multipartite reduced density matrices. For pure
three-qubit states, we show an trade-off relation between the
coherence distribution and the three tangle of quantum entanglement.

\section{Trade-off relations of multi-qubit coherence}

The $l_{1}$ norm quantum coherence of any quantum state $\rho=\sum
\rho_{ij}|i\rangle\langle j|$ is given by the non-diagonal entries of $\rho$ \cite{tbmc},
\begin{eqnarray}
C_{l_{1}}(\rho)=\sum_{i,j,i\neq j}|\rho_{ij}|.
\end{eqnarray}
In the following, we denote for a tripartite state $\rho_{ABC}$,
$C_{123}=C_{l_{1}}(\rho_{ABC})$, $C_{12}=C_{l_{1}}(\rho_{AB})$,
$C_{13}=C_{l_{1}}(\rho_{AC})$, $C_{23}=C_{l_{1}}(\rho_{BC})$, where
$\rho_{AB}=Tr_{C}(\rho_{ABC})$, $\rho_{AC}=Tr_{B}(\rho_{ABC})$ and $\rho_{BC}=Tr_{A}(\rho_{ABC})$ are
the reduced density matrices.

 According to the definition of $l_{1}$ norm quantum coherence, for any $d$-dimensional quantum state $\rho$, one has $C_{l_{1}}(\rho)\leq d-1$,  where the upper bound is attained for maximally coherent states,
 it is obvious that the coherence of each subsystem
is not greater than the coherence of this quantum state. For
example, $C_{123}\geq C_{12}$ , $C_{123}\geq C_{13}$, $C_{123}\geq
C_{1}$.
 In order to study the trade-off relation between quantum states and its subsystems, we first consider three-qubit case.

\begin{theorem}
For any three-qubit quantum state $\rho_{ABC}=\sum_{i,j,k,i',j',k'=0}^{1}\rho_{ijk}^{i'j'k'}|ijk\rangle\langle i'j'k'|$, we have
\begin{eqnarray}\label{Thm1}
C_{123}\geq\frac{C_{12}+C_{13}+C_{23}}{2}.
\end{eqnarray}
\end{theorem}

{\bf{Proof:}} From
$$
C_{123}={\sum_{i,j,k=0}^{1}\sum_{i',j',k'=0}^{1}}_{i\neq i' or j\neq j' or k\neq
k'}|\rho_{ijk}^{i'j'k'}|,
$$
by using the triangle inequality $|a|+|b|\geq|a+b|$, we have
\begin{eqnarray*}
& &2C_{123}\nonumber\\
&\geq&{\sum_{i,j=0}^{1}\sum_{i',j'=0}^{1}}_{i\neq i' or j\neq j'}|\sum_{k=0}^{1}\rho_{ijk}^{i'j'k}|\nonumber\\
& &+{\sum_{i,k=0}^{1}\sum_{i',k'=0}^{1}}_{i\neq i' or k\neq k'}|\sum_{j=0}^{1}\rho_{ijk}^{i'jk'}|\nonumber\\
& &+{\sum_{j,k=0}^{1}\sum_{j',k'=0}^{1}}_{j\neq j' or k\neq k'}|\sum_{i=0}^{1}\rho_{ijk}^{ij'k'}|\nonumber\\
& &+D\\
&=&C_{12}+C_{13}+C_{23}+D,
\end{eqnarray*}
where
$D=|\rho_{000}^{011}|+|\rho_{000}^{101}|+|\rho_{000}^{110}|+|\rho_{001}^{010}|+|\rho_{001}^{100}|+|\rho_{001}^{111}|+|\rho_{010}^{100}|
+|\rho_{010}^{111}|+|\rho_{011}^{101}|+|\rho_{011}^{110}|+|\rho_{100}^{111}|+|\rho_{101}^{110}|+|\rho_{011}^{000}|+|\rho_{101}^{000}|
+|\rho_{110}^{000}|+|\rho_{010}^{001}|+|\rho_{100}^{001}|+|\rho_{111}^{001}|+|\rho_{100}^{010}|+|\rho_{111}^{010}|+|\rho_{101}^{011}|
+|\rho_{110}^{011}|+|\rho_{111}^{100}|+|\rho_{110}^{101}|+2(|\rho_{000}^{111}|+|\rho_{001}^{110}|+|\rho_{010}^{101}|+|\rho_{011}^{100}|
+|\rho_{100}^{011}|+|\rho_{101}^{010}|+|\rho_{110}^{001}|+|\rho_{111}^{000}|)$,
which gives rise to (\ref{Thm1}). \hfill \rule{1ex}{1ex}

For example, for a pure incoherent product state $|\psi\rangle=a_{ijk}|ijk\rangle$, one has trivially $C_{123}=C_{12}=C_{13}=C_{23}=0$. For coherent state of the form,
$|\psi\rangle=a_{000}|000\rangle+a_{100}|100\rangle$, we get
$C_{123}=C_{12}=C_{13}=|a_{000}a_{100}^{*}|+|a_{100}a_{000}^{*}|$ and $C_{23}=0$. The equality holds in this case, $C_{123}=(C_{12}+C_{13}+C_{23})/{2}$, which
gives rise to $C_{123}\leq C_{12}+C_{13}$ as $C_{23}=0$. Therefore,
the conjecture in \cite{ewaa}, $C_{123}\geq C_{12}+C_{13}$ is invalid in this case.

In \cite{ewaa}, the author discussed that the trade-off relation $C_r(\rho_{ABC})\geq
C_r(\rho_{AB})+C_r(\rho_{AC})$ does not hold by the relative entropy. Similarly, we also give a class of quantum states violating the
trade-off relation $C_{123}\geq C_{12}+C_{13}$.

Due to correlated coherence $C_{cc}(\rho_{AB})$ is always nonnegative \cite{awwa}, we have $C_{l_{1}}(\rho_{AB})\geq C_{l_{1}}(\rho_{A})+C_{l_{1}}(\rho_{B})$, similarly, the trade-off relations $C_{l_{1}}(\rho_{AC})\geq C_{l_{1}}(\rho_{A})+C_{l_{1}}(\rho_{C})$  and $C_{l_{1}}(\rho_{BC})\geq C_{l_{1}}(\rho_{B})+C_{l_{1}}(\rho_{C})$ are hold, i.e, $C_{12}\geq C_{1}+C_{2}$ ,$C_{13}\geq C_{1}+C_{3}$ and $C_{23}\geq C_{2}+C_{3}$. Therefore, we have
\begin{eqnarray}\label{ccaa}
C_{123}&\geq&\frac{C_{12}+C_{13}+C_{23}}{2}\nonumber\\
&\geq&C_{1}+C_{2}+C_{3}
\end{eqnarray}
For the trade-off relation $C_{123}\geq C_{12}+C_{13}$, we have
\begin{eqnarray}\label{jokj}
C_{123}&\geq& C_{12}+C_{13}\nonumber\\
&\geq&2C_{1}+C_{2}+C_{3}
\end{eqnarray}
When $2C_{1}\geq C_{123}$, the  inequality (\ref{jokj}) does not hold. Similarly, when $2C_r(\rho_{A})\geq C_r(\rho_{ABC})$, the trade-off relation $C_r(\rho_{ABC})\geq C_r(\rho_{AB})+C_r(\rho_{AC})$ is invalid. Therefore, we give the trade-off relation  between tripartite coherence and  bipartite coherence in Theorem 1. In addition, we can get the trade-off relation by using the triangle inequality as follows.
\begin{eqnarray}
C_{123}\geq C_{1}+C_{23}
\end{eqnarray}
where $C_{1}={\sum_{i=0}^{1}\sum_{i'=0}^{1}}_{i\neq i' }|\sum_{j=0}^{1}\sum_{k=0}^{1}\rho_{ijk}^{i'jk}|$, similarly, we have $C_{123}\geq C_{2}+C_{13}$ and $C_{123}\geq C_{3}+C_{12}$.

Generalizing Theorem 1 to n-qubit case, we have, see proof in Appendix,

\begin{theorem}
For any n-qubit quantum state $\rho=\sum\rho_{i_{1}i_{2}\cdots i_{n}}^{j_{1}j_{2}\cdots j_{n}}|i_{1}i_{2}\cdots i_{n}\rangle\langle j_{1}j_{2}\cdots j_{n}|,$ we have
\begin{eqnarray}\label{daa}
&&C_{123\cdots n}\geq\nonumber\\[2mm]
&&\frac{C_{123\cdots
(n-1)}+C_{123\cdots(n-2)n}+\cdots+C_{234\cdots(n-1)n}}{n-1}.
\end{eqnarray}
\end{theorem}

The lower bound of (\ref{daa}) can be further expressed as the coherence of
all m-partite reduced states of the n-qubit state. Let
$\Gamma(m,n)=\{a_{1}a_{2}\cdots a_{m}|1\leq
a_{1}<a_{2}<\cdots<a_{m}\leq n\}$ denote the set of m different
elements from n. For example, $\Gamma(2,4)=\{12,13,14,23,24,34\}$.
Then for any given $m$, we have, see proof in Appendix,

\begin{corollary}
For any n-qubit quantum state $\rho=\sum\rho_{i_{1}i_{2}\cdots
i_{n}}^{j_{1}j_{2}\cdots j_{n}}|i_{1}i_{2}\cdots i_{n}\rangle\langle
j_{1}j_{2}\cdots j_{n}|$, we have
\begin{eqnarray}\label{C1}
C_{123\cdots
n}\geq\frac{\sum_{a\in\Gamma(m,n)}C_{a}}{C_{n-1}^{m-1}},
\end{eqnarray}
where the combination $C_{n-1}^{m-1}$ represents the maximum number of
occurrences of the element $\rho_{i_{1}i_{2}\cdots
i_{n}}^{j_{1}j_{2}\cdots j_{n}}$ on the right side of the inequality.
\end{corollary}

In particular, taking $a\in\Gamma(2,3)$ or
$a\in\Gamma(n-1,n)$, one gets the Theorem 1 or Theorem 2,
respectively.

The results in Corollary 1 can be straightforwardly generalized to multi-qudit case.

\begin{corollary}
For any $n$-qudit $\rho=\sum_{i_{1},i_{2},\cdots,i_{n},j_{1},j_{2},\cdots,j_{n}=0}^{d-1}\rho_{i_{1}i_{2}\cdots i_{n}}^{j_{1}j_{2}\cdots j_{n}}|i_{1}i_{2}\cdots i_{n}\rangle\langle j_{1}j_{2}\cdots j_{n}|,$ we have
\begin{eqnarray}\label{Thm2}
C_{123\cdots
n}\geq\frac{\sum_{a\in\Gamma(m,n)}C_{a}}{C_{n-1}^{m-1}}.
\end{eqnarray}
\end{corollary}

(\ref{Thm2}) can be proved by similar derivations to the proof of Theorem 2 and Corollary 1. In fact,
it is valid for any multipartite states with different individual dimensions.

Above results valid for any mixed quantum states.
Next, we consider the relationship between the coherence and the
entanglement for the 3-qubit pure states.

\begin{theorem}
For any three-qubit pure state
$|\psi\rangle_{ABC}=\sum_{i,j,k=0}^{1}a_{ijk}|ijk\rangle$, we have
\begin{eqnarray}\label{Thm3}
C_{123}\geq\frac{C_{12}+C_{13}+C_{23}}{2}+\tau_{123},
\end{eqnarray}
where $\tau_{123}=4|d_{1}-2d_{2}+4d_{3}|$ is entanglement tangle \cite{vcjk},
$d_{1}=a_{000}^{2}a_{111}^{2}+a_{001}^{2}a_{110}^{2}+a_{010}^{2}a_{101}^{2}+a_{100}^{2}a_{011}^{2}$,
$d_{2}=a_{000}a_{111}a_{011}a_{100}+a_{000}a_{111}a_{101}a_{010}
+a_{000}a_{111}a_{110}a_{001}+a_{011}a_{100}a_{101}a_{010}
+a_{011}a_{100}a_{110}a_{001}+a_{101}a_{010}a_{110}a_{001}$ and
$d_{3}=a_{000}a_{110}a_{101}a_{011}+a_{111}a_{001}a_{010}a_{100}$.
\end{theorem}

In \cite{crmp}, the author proposes a trade-off upper bound for the tripartite system , $C_{123} \leq C_{1}+C_{2}+C_{3}+C_{1:2:3}$,
where $C_{1:2:3}$ is an intrinsic coherence, defined as minimized over the set of separable states on the
tripartite system. For the three-qubit pure state, we give a lower bound inequality that combines  entanglement
measure in Theorem 3. According to the inequality (\ref{ccaa}), we have
\begin{eqnarray}
C_{123}\geq C_{1}+C_{2}+C_{3}+\tau_{123}.
\end{eqnarray}

As examples, let us consider the GHZ state and the W state. For the
GHZ state $|GHZ\rangle=\cos\phi|000\rangle+\sin\phi|111\rangle$,
$\phi\in[0,2\pi),$ we have $C_{123}=2|\sin\phi \cos\phi|$,
$C_{12}=C_{13}=C_{23}=0$, $\tau_{123}=4|\cos^{2}\phi \sin^{2}\phi|.$
As $2|\sin\phi \cos\phi|\geq 4|\cos^{2}\phi \sin^{2}\phi|,$ one gets
the inequality (\ref{Thm3}). For the W state
$|W\rangle=\sin\theta\cos\phi|100\rangle+\sin\theta
\sin\phi|010\rangle+\cos\theta|001\rangle$ with $0\leq\phi<2\pi$ and
$0\leq\theta<\pi,$ we get $C_{123}=2(|\sin^{2}\theta \sin\phi
\cos\phi|+|\sin\theta \cos\theta \cos\phi|+|\sin\theta \cos\theta
\sin\phi|),C_{12}=2|\sin^{2}\theta \sin\phi \cos\phi|$,
$C_{13}=2|\sin\theta \cos\theta \cos\phi|$, $C_{23}=2|\sin\theta
\cos\theta \sin\phi|$ and $\tau_{123}=0$, which satisfy the
inequality (\ref{Thm3}). This example shows that both GHZ and W
states obey the same inequality (\ref{Thm3}), which is different from the case in
\cite{crmp}, where the GHZ and W states satisfy different inequalities under the relative entropy of coherence.

\section{conclusion and discussion}

We have studied the trade-off relations under the $l_{1}$-norm of
quantum coherence. The general trade-off relations satisfied by the
coherence of multipartite quantum states have been derived. For pure
three-qubit case, it has been explicitly shown that the trade-off
relation is lower bounded by the three tangle of quantum entanglement.
These results may highlight further studies on coherence and correlation
distributions in multipartite quantum systems.

We have proved that the trade-off relation $C(\rho_{ABC})\geq
C(\rho_{AB})+C(\rho_{AC})$ is invalid by the method of relative
entropy and $l_{1}$ norm. Here, we given the definition of
correlated coherence by the $l_{1}$norm \cite{awwa}, it is obvious
that one also can define the correlated coherence by the entopic
coherence measure and the convex roof coherence measure. So we
believe that the inequality (\ref{Thm1}) holds in other quantum
coherence measures.

\bigskip
Acknowledgments: We thank anonymous reviewers for their suggestions
for improving the article. This work is supported by the NSF of
China under Grant Nos. 11861031 and 11675113, and Beijing Municipal
Commission of Education (KM201810011009).

\section{appendix}
\subsection{Proof of Theorem 2}
{\bf{Proof:}} From the definitions
\begin{eqnarray*}
&&C_{123\cdots n}=\\
&&\sum_{\substack{i_{1},i_{2},\cdots,i_{n}=0 \\ {i_{1}\neq j_{1} \
or \ i_{2}\neq
j_{2}}}}^{1}\sum_{\substack{j_{1},j_{2},\cdots,j_{n}=0
\\ { or\cdots or \ i_{n}\neq j_{n}}}}^{1}|\rho^{j_{1}j_{2}\cdots j_{n}}_{i_{1}i_{2}\cdots i_{n}}| \ ,\\[2mm]
&&C_{123\cdots (n-1)}=\\
&&\sum_{\substack{i_{1},i_{2},\cdots,i_{n-1}=0 \\ {i_{1}\neq j_{1} \
or \ i_{2}\neq
j_{2}}}}^{1}\sum_{\substack{j_{1},j_{2},\cdots,j_{n-1}=0
\\ {or\cdots or \ i_{n-1}\neq j_{n-1}}}}^{1}|\sum_{i_{n}=0}^{1}\rho^{j_{1}j_{2}\cdots j_{n-1}i_{n}}_{i_{1}i_{2}\cdots i_{n-1}i_{n}}| \ ,\\[2mm]
&&C_{123\cdots (n-2)n}=\\
&&\sum_{\substack{i_{1},\cdots,i_{n-2},i_{n}=0 \\ {i_{1}\neq j_{1}
or\cdots or }}}^{1}\sum_{\substack{j_{1},\cdots,j_{n-2},j_{n}=0
\\ { \ i_{n-2}\neq j_{n-2}or i_{n}\neq j_{n} }}}^{1}|\sum_{i_{n-1}=0}^{1}\rho^{j_{1}\cdots j_{n-2}i_{n-1}j_{n}}_{i_{1}\cdots i_{n-2}i_{n-1}i_{n}}| \ ,\\
&&\vdots\\
&&C_{234\cdots n}=\\
&&\sum_{\substack{i_{2},i_{3},\cdots,i_{n}=0 \\ {i_{2}\neq j_{2} \
or \ i_{3}\neq
j_{3}}}}^{1}\sum_{\substack{j_{2},j_{3},\cdots,j_{n}=0
\\ {or\cdots or \ i_{n}\neq j_{n}}}}^{1}|\sum_{i_{1}=0}^{1}\rho^{i_{1}j_{2}\cdots j_{n}}_{i_{1}i_{2}\cdots i_{n}}|,
\end{eqnarray*}
similar to the proof of Theorem 1, by using triangular inequalities
and taking into account the number of times of the same element
appearing on both sides of the inequalities, we obtain (3). \hfill
\rule{1ex}{1ex}
\subsection{Proof of Corollary 1}
{\bf{Proof:}}  According to Theorem 2, we get
\begin{eqnarray*}
& &C_{123\cdots n}\nonumber\\[2mm]
&\geq&\frac{C_{123\cdots(n-1)}+C_{123\cdots(n-2)n}+\cdots+C_{234\cdots(n-1)n}}{n-1}\\
&=&\frac{\sum_{a\in\Gamma(n-1,n)}C_{a}}{C_{n-1}^{n-2}}.
\end{eqnarray*}
Hence
\begin{eqnarray*}
& &C_{123\cdots n}\nonumber\\[2mm]
&\geq&\frac{C_{123\cdots
(n-1)}+C_{123\cdots(n-2)n}+\cdots+C_{234\cdots(n-1)n}}{n-1}.\\
&\geq&{\frac{C_{123\cdots(n-2)}+C_{12\cdots(n-3)(n-1)}+\cdots+C_{234\cdots(n-1)}}{(n-2)(n-1)}}\\
& &+{\frac{C_{123\cdots(n-2)}+C_{12\cdots(n-3)n}+\cdots+C_{234\cdots(n-2)n}}{(n-2)(n-1)}}\\
& &+\cdots\\[2mm]
& &+{\frac{C_{234\cdots(n-1)}+C_{23\cdots(n-2)n}+\cdots+C_{345\cdots n}}{(n-2)(n-1)}}\\
&=&\frac{C_{123\cdots(n-2)}+C_{12\cdots(n-3)(n-1)}+\cdots+C_{345\cdots
n}}{\frac{(n-1)(n-2)}{2}}.
\end{eqnarray*}
According to the characteristics of the combination, each element
appears twice on the right side of the second inequality. One has
\begin{eqnarray*}
C_{123\cdots
n}\geq\frac{\sum_{a\in\Gamma(n-2,n)}C_{a}}{C_{n-1}^{n-3}}.
\end{eqnarray*}

Similarly, we obtain
\begin{eqnarray*}
C_{123\cdots n}&\geq&\frac{\sum_{a\in\Gamma(n-3,n)}C_{a}}{C_{n-1}^{n-4}}.\\
C_{123\cdots n}&\geq&\frac{\sum_{a\in\Gamma(n-4,n)}C_{a}}{C_{n-1}^{n-5}}.\\
&\vdots&\\
C_{123\cdots
n}&\geq&\frac{\sum_{a\in\Gamma(1,n)}C_{a}}{C_{n-1}^{0}},
\end{eqnarray*}
which give rise to (\ref{C1}). \hfill \rule{1ex}{1ex}

\subsection{Proof of Theorem 3}
{\bf{Proof:}}  The two-qubit reduced density matrices of $\rho_{ABC}=|\psi\rangle_{ABC}\langle\psi|$ are given by
$$\rho_{AB}=\sum_{i,j=0}^{1}\sum_{i',j'=0}^{1}\sum_{k=0}^{1}a_{ijk}a_{i'j'k}^{*}|ij\rangle\langle i'j'|,$$
$$\rho_{AC}=\sum_{i,k=0}^{1}\sum_{i',k'=0}^{1}\sum_{j=0}^{1}a_{ijk}a_{i'jk'}^{*}|ik\rangle\langle  i'k'|,$$
$$\rho_{BC}=\sum_{j,k=0}^{1}\sum_{j',k'=0}^{1}\sum_{i=0}^{1}a_{ijk}a_{ij'k'}^{*}|jk\rangle\langle  j'k'|.$$
According to the proof of Theorem 1 and the fact that $|xy^{*}|=|xy|$
for any complex numbers x and y, we have
\begin{eqnarray*}
C_{123}\geq\frac{C_{12}+C_{13}+C_{23}}{2}+\frac{D'}{2},
\end{eqnarray*}
where
\begin{eqnarray*}
\frac{D'}{2}&=&|a_{000}a_{011}|+|a_{000}a_{101}|+|a_{000}a_{110}|+|a_{001}a_{010}|\nonumber\\
& &+|a_{001}a_{100}|+|a_{001}a_{111}|+|a_{010}a_{100}|+|a_{010}a_{111}|\nonumber\\
& &+|a_{011}a_{110}|+|a_{011}a_{101}|+|a_{100}a_{111}|+|a_{101}a_{110}|\nonumber\\
& &+2(|a_{000}a_{111}|+|a_{001}a_{110}|+|a_{010}a_{101}|+|a_{011}a_{100}|).\\
\end{eqnarray*}

From the inequality $a^{2}+b^{2}\geq2ab$ for $a\geq0$ and $b\geq0$, we have
\begin{eqnarray*}
1&=&\sum_{i,j,k=0}^{1}|a_{ijk}|^{2}\nonumber\\
&\geq&2(|a_{000}a_{111}|+|a_{001}a_{110}|+|a_{010}a_{101}|+|a_{100}a_{011}|)\\
&\geq& 0.
\end{eqnarray*}
Hence
\begin{eqnarray*}
& &2(|a_{000}a_{111}|+|a_{001}a_{110}|+|a_{010}a_{101}|+|a_{100}a_{011}|)\\
&\geq& [2(|a_{000}a_{111}|+|a_{001}a_{110}|+|a_{010}a_{101}|+|a_{100}a_{011}|)]^{2}\\
&=&4[|a_{000}|^{2}|a_{111}|^{2}+|a_{001}|^{2}|a_{110}|^{2}\\
& &+|a_{010}|^{2}|a_{101}|^{2}+|a_{100}|^{2}|a_{011}|^{2}\\
& &+2(|a_{000}a_{111}a_{011}a_{100}|+|a_{000}a_{111}a_{101}a_{010}|\\
& &+|a_{000}a_{111}a_{110}a_{001}|+|a_{011}a_{100}a_{101}a_{010}|\\
& &+|a_{011}a_{100}a_{110}a_{001}|+|a_{101}a_{010}a_{110}a_{001}|)]\\
&\geq& 4|d_{1}-2d_{2}|.
\end{eqnarray*}
Similarly
\begin{eqnarray*}
1&\geq&|a_{000}|^{2}+|a_{011}|^{2}+|a_{101}|^{2}+|a_{110}|^{2}\\
&\geq& 2(|a_{000}a_{011}|+|a_{101}a_{110}|) \\
&\geq& 0.
\end{eqnarray*}
Therefore,
\begin{eqnarray*}
& &|a_{000}a_{011}|+|a_{101}a_{110}|\\
&\geq&2(|a_{000}a_{011}|+|a_{101}a_{110}|)^{2}\\
&=&2(|a_{000}a_{011}|^{2}+|a_{101}a_{110}|^{2}+2|a_{000}a_{110}a_{101}a_{011}|)\\
&\geq&8|a_{000}a_{110}a_{101}a_{011}|.
\end{eqnarray*}
And similarly,
\begin{eqnarray*}
|a_{000}a_{101}|+|a_{011}a_{110}|\geq8|a_{000}a_{110}a_{101}a_{011}|,\\
|a_{000}a_{110}|+|a_{011}a_{101}|\geq8|a_{000}a_{110}a_{101}a_{011}|,\\
|a_{001}a_{010}|+|a_{100}a_{111}|\geq8|a_{111}a_{001}a_{010}a_{100}|,\\
|a_{001}a_{100}|+|a_{010}a_{111}|\geq8|a_{111}a_{001}a_{010}a_{100}|,\\
|a_{001}a_{111}|+|a_{010}a_{100}|\geq8|a_{111}a_{001}a_{010}a_{100}|.
\end{eqnarray*}
Thus we have
\begin{eqnarray*}
& &|a_{000}a_{011}|+|a_{000}a_{101}|+|a_{000}a_{110}|+|a_{001}a_{010}|\\
& &+|a_{001}a_{100}|+|a_{001}a_{111}|+|a_{010}a_{100}|+|a_{010}a_{111}|\\
& &+|a_{011}a_{110}|+|a_{011}a_{101}|+|a_{100}a_{111}|+|a_{101}a_{110}|\\
&\geq&16(|a_{000}a_{110}a_{101}a_{011}|+|a_{111}a_{001}a_{010}a_{100}|)\\
&\geq&4|4d_{3}|,
\end{eqnarray*}
Therefore, we obtain
\begin{eqnarray*}
\frac{D'}{2}\geq4|d_{1}-2d_{2}+4d_{3}|=\tau_{123},
\end{eqnarray*}
and then (\ref{Thm3}).
\hfill \rule{1ex}{1ex}

\end{document}